\documentclass[lettersize,journal, twoside]{IEEEtran}
\usepackage{amsmath,amssymb,amsfonts}
\usepackage{algorithmic}
\usepackage{algorithm}
\usepackage{array}
\usepackage{textcomp}
\usepackage{stfloats}
\usepackage{url}
\usepackage{verbatim}
\usepackage{graphicx}
\usepackage{cite}
\usepackage{mathrsfs}
\usepackage{mathtools}
\usepackage{physics}
\usepackage{subcaption}
\usepackage{etoolbox}
\usepackage{bm}

\newcommand{\Z}{\mathbb{Z}}
\newcommand{\R}{\mathbb{R}}
\newcommand{\G}{\mathbb{G}}

\newcommand{\M}{\mathcal{M}}
\newcommand{\C}{\mathcal{C}}
\newcommand{\E}{\mathcal{E}}
\newcommand{\Gen}{\mathsf{Gen}}
\newcommand{\Enc}{\mathsf{Enc}}
\newcommand{\Dec}{\mathsf{Dec}}
\newcommand{\Eval}{\mathsf{Eval}}
\newcommand{\Ecd}{\mathsf{Ecd}}
\newcommand{\Dcd}{\mathsf{Dcd}}

\newcommand{\pk}{\mathsf{pk}}
\newcommand{\skd}{\mathsf{sk_d}}
\newcommand{\skh}{\mathsf{sk_h}}

\newtheorem{remark}{Remark}

\begin{document}

\title{Malleability-Resistant Encrypted Control System with Disturbance Compensation and Real-Time Attack Detection}


\author{
Naoki Aizawa,~\IEEEmembership{Student Member,~IEEE,}
Keita Emura,~and 
Kiminao Kogiso,~\IEEEmembership{Member, IEEE}
\thanks{This work was supported in part by JSPS KAKENHI JP22H01509 and JP23K22779.}
\thanks{Naoki Aizawa is with the Department of Mechanical and Intelligent Systems Engineering, The University of Electro-Communications, Chofu, Tokyo 1828585, Japan. Corresponding author(e-mail: aizawanaoki@uec.ac.jp).}
\thanks{Keita Emura is with Institute of Science and Engineering, Kanazawa University, Kanazawa, Ishikawa 920–1192, Japan. }
\thanks{Kiminao Kogiso is with the Department of Mechanical and Intelligent Systems Engineering, The University of Electro-Communications, Chofu, Tokyo 1828585, Japan.}
}






\maketitle

\begin{abstract}
This study proposes an encrypted PID control system with a disturbance observer (DOB) using a keyed-homomorphic encryption (KHE) scheme, aiming to achieve control performance while providing resistance to malleability-based attacks. 
The controller integrates a DOB with a PID structure to compensate for modeling uncertainties by estimating and canceling external disturbances. To enhance security, the system is designed to output error symbols when ciphertexts are falsified during decryption or evaluation, enabling real-time detection of malleability-based signal or parameter falsification.
To validate the proposed method, we conduct stage positioning control experiments and attack detection tests using an industrial linear stage. The results show that the encrypted DOB-based PID controller outperforms a conventional encrypted PID controller in terms of tracking accuracy. Furthermore, the system successfully detects two types of malleability-based attacks: one that destabilizes the control system, and another that degrades its performance.
The primary contributions of this study are: (i) the implementation of a KHE-based encrypted DOB-PID controller, (ii) the improvement of control performance under uncertainties, and (iii) the experimental demonstration of attack detection capabilities in encrypted control systems.
\end{abstract}


\section{Introduction}
The cybersecurity of networked control systems is critical.
As these systems continue to evolve, the risk of cyberattacks targeting them has increased.
For example, Stuxnet destroyed centrifuges at the Iranian nuclear facility~\cite{Lan11}, and a false data injection attack compromised sensor measurements in Ukrainian power grid~\cite{Liang16}.
Among various attacks on control systems~\cite{Mo09,Mo10,Cetinkaya19,Pasqualetti13}, eavesdropping attacks not only steal communication signals but can also act as a precursor to more destructive intrusions~\cite{Chong19}.
Other attacks involve stealing operational information, such as controller parameters, by hacking into control devices.
Therefore, enhancing the security of control systems is essential to protect them from cyber threats.

A promising approach to enhancing the security of networked control systems is encrypted control~\cite{Kogiso15, Kim22, Darup21, Darup23, Bian25}, which protects control system information by employing homomorphic encryption schemes.
This enables the direct computation of encrypted control inputs from encrypted measurements, without requiring decryption.
Encrypted control relies on various types of homomorphic encryption schemes, such as multiplicative homomorphic encryption~\cite{Rivest78, ElGamal85}, additive homomorphic encryption~\cite{Paillier99}, fully homomorphic encryption~\cite{Brakerski14, Gentry09, Cheon17}, and somewhat homomorphic encryption~\cite{Dyer19}.
Furthermore, encryption schemes specifically adapted for control systems have been developed, including resilient homomorphic encryption\cite{Fauser21, Fauser24} and dynamic-key homomorphic encryption\cite{Teranishi22}.
Although these methods can be effective for eavesdropping and mitigate several types of attack risks, they remain vulnerable to malleability, which is an inherent property of homomorphic encryption schemes~\cite{Katz09}.
Malleability is a security vulnerability that allows arithmetic operations to be performed directly on ciphertexts without knowledge of the encryption keys.
In malleability-based attacks on encrypted control systems, adversaries can manipulate encrypted sensor measurements, control parameters, or control inputs without decryption, thereby destabilizing the control system~\cite{Teranishi19} and compromizing the behavior of the plant~\cite{Kwon25,Kwon25journal,Blevins25}.

The design of encrypted control systems must ensure resistance to malleability while maintaining satisfactory control performance across a range of plant dynamics.
To address the vulnerability associated with malleability, keyed-homomorphic encryption (KHE) schemes have proven useful, as proposed in~\cite{Emu13,Lib14,Jut15, Mae22,Shi24}.
The concept of KHE introduces an additional private key dedicated to homomorphic operations, enabling the detection of signal or parameter falsification that exploits the malleability.
To date, the only appliaction of KHE to control systems is the encrypted PID control system developed in~\cite{Miy23}.
Although this approach is effective in integrating encryption into control, it is not sufficient for ensuring satisfactory performance, particularly in systems with friction or modeling uncertainties, where PID control may fall short.
To overcome the limitations of PID control in dealing with uncertainties, disturbance observers (DOBs)\cite{Chen15} are widely used in industrial systems.
Several encrypted control systems incorporating DOBs have been proposed\cite{Teranishi19dob,Takanashi25}.
However, all of these are based on the ElGamal encryption scheme, which is vulnerable to malleability-based attacks, thereby limiting their security guarantees.
Furthermore, since numerical simulations cannot adequately account for uncertainties, it is essential to conduct experiments that not only evaluate the impact of quantization errors and determine appropriate encryption parameters, but also verify the system’s ability to detect attacks in scenarios where an adversary exploits the malleability of the encryption scheme to compromise the control device and destabilize the system.

This study aims to propose an encrypted DOB-based PID control system using a KHE scheme, in order to achieve both control performance and resistance to malleability-based attacks.
To this end, a disturbance observer (DOB) is integrated with a PID controller so that the estimated disturbnace can be added to the control input, thereby compensating for modeling uncertainties.
The proposed system is also designed to output error symbols when ciphertexts are falsified during decryption or evaluation, enabling the detection of malleability-based attacks.
Furthermore, to validate the effectiveness of the proposed method, stage positioning control experiments and attack detection tests are conducted using an industrial linear stage.
The experimental results demonstrate that the proposed encrypted DOB-based PID control system outperforms a conventional encrypted PID controller in terms of tracking performance.
For attack detection, two scenarios of malleability-based control parameter falsification are evaluated: one that destabilizes the system, similar to the Stuxnet attack, and another that degrades performance.
The results confirm that the system can detect and localize attacked components in real time.

The contributions of this study are threefold:
i) the practical construction of an encrypted DOB-based PID controller using a KHE scheme, 
ii) the demonstration of improved control performance under uncertainty, and 
iii) the successful detection of malleability-based attacks through experimental validation.
Unlike the method proposed in~\cite{Miy23}, this study integrates a disturbance observer into the control structure and addresses attack scenarios that target system destabilization.
The remainder of this paper is organized as follows.
Section~\ref{sec:pre} introduces the notations and KHE scheme as preliminaries.
Section~\ref{sec:prob} formulates the proposed encrypted DOB-based PID controller.
Section~\ref{sec:ev} presents experimental results evaluating the tracking control performance.
Section~\ref{sec:ad} demonstrates cyberattack tests to validate the feasibility of real-time detection.
Finally, Section~\ref{sec:con} concludes the paper.

\section{Preliminaries}\label{sec:pre}
This section defines the notations of variables, functions, and a quantizer, and introduces KHE for encrypting the controller and facilitating real-time attack detection.

\subsection{Notations}
The sets of real numbers, integers, plaintext spaces, and ciphertext spaces are denoted by $\R$, $\Z$, $\M$, and $\C$, respectively.
We define
$\R^{+}:=\{x\in\R\mid 0<x\}$, 
$\Z^{+}:=\{z\in\Z\mid 0<z\}$,
$\Z^+_0:=\{z\in\Z\mid 0\leq z\}$,
$\Z_n:=\{z\in\Z\mid0 \leq z < n\}$,
$\Z_n^{+}:=\{z\in\Z\mid 0 < z < n\}$, and
$\mathfrak{P}_{a}^{b}:=\{a^{i}\bmod b \,|\, i \in \Z_b\}$.
A multiplicative cyclic group is defined as $\G:=\{g^i \bmod p \mid i \in \Z_q \}$ such that $g^q \bmod p=1$ and $p-1 \bmod q =0$ with generator $g$ of the cyclic group $\G$.
A set of vectors of size $n$ is denoted by $\R^{n}$.
The $j$th element of vector $v$ is denoted by $v_j$. 
The set of matrices of size $m \times n$ is denoted by $\mathbb{R}^{m\times n}$.
The $(i,j)$ entry of matrix $M$ is denoted by $M_{ij}$.
$\textbf{0}$ denotes a zero vector or zero matrix of an appropriate dimension.
The greatest common divisor of the two positive integers $a$, $b\in\mathbb{Z}^+$ is denoted by $\textrm{gcd}(a,b)$. 
The minimal residue of integer $a\in\mathbb{Z}$ modulo $m\in\mathbb{Z}^{+}$ is defined as
$a\,\text{Mod}\,m=b$ if $b<|b - m|$ holds; otherwise, $a\,\text{Mod}\,m=b-m$, where $b=a\bmod m$.

Let $p$ be an odd prime number and $z$ be an integer satisfying $\text{gcd}(z,p)=1$.
If there exists an integer $b$ such that $b^2=z\bmod p$, then integer $z$ is a quadratic residue modulo $p$.
If such an integer $b$ does not exist, then $z$ is a quadratic nonresidue modulo $p$.
This can be expressed using the Legendre symbol $(\cdot/\cdot)_L$ as follows:
$\left({z}/{p}\right)_{L}= z^{\frac{p-1}{2}}\, \text{Mod}\,p=1$ if $z$ is a quadratic residue; otherwise, -1.
The rounding function $\lceil\cdot\rfloor$ of $\sigma\in\mathbb{R}^{+}$ to the nearest positive integer is defined as 
$\lceil\sigma\rfloor=\lfloor\sigma+0.5\rfloor$ if $\sigma \geq 0.5$; otherwise, $\lceil\sigma\rfloor=1$, where $\lfloor \cdot\rfloor$ denotes the floor function.

This study uses a quantizer that maps $x\in\R$ onto $\bar{x}:=(\bar{x}^1,\bar{x}^2)$, proposed in \cite{Miy23}.
An encoding map $\Ecd_{\gamma}:=\mathscr{C}\circ\mathscr{A}_{\gamma}$ and a decoding map $\Dcd_{\gamma}:=\mathscr{B}_{\gamma}\circ\mathscr{D}$ are described as follows:
\begin{align*}
\mathscr{A}_{\gamma}&:\mathbb{R}\rightarrow\mathfrak{P}^{q}_{2}\times\mathbb{Z}_q^{+},\\
  &:x\mapsto
  \left\{
    \begin{array}{ll}
      \left(1, \lceil \gamma|x| \rfloor\bmod q\right) & \text{if}\ \,x \geq 0,\\
      \left(2, \lceil \gamma|x| \rfloor\bmod q\right) & \text{if}\ \,x < 0,
    \end{array}
  \right.\\
  \mathscr{B}_{\gamma}&:\mathfrak{P}^{q}_{2}\times\mathbb{Z}_q^{+}\rightarrow\mathbb{R},\\
  &:(\zeta, z)\mapsto\left(\frac{\zeta}{3}\right)_{L}\frac{z}{\gamma}:=\check{x},\\
  \mathscr{C}&:\mathfrak{P}^{q}_{2}\times\mathbb{Z}_q^{+}\rightarrow\mathbb{G}^2,\\
  &:(\zeta, z)\mapsto\left(\left(\frac{\zeta}{p}\right)_{L}\zeta, \left(\frac{z}{p}\right)_Lz\right)\bmod p:=(\bar{x}^1,\bar{x}^2),\\
  \mathscr{D}&:\mathbb{G}^2\rightarrow\mathfrak{P}^{q}_{2}\times\mathbb{Z}_q^{+},\\
  &:(\bar{x}^1,\bar{x}^2)\mapsto (|\bar{x}^1 \, \text{Mod} \, p|,|\bar{x}^2 \, \text{Mod} \, p|),
\end{align*}
where $\gamma\in\R^{+}$ is a quantization gain, $\zeta\in\{1,2\}$ and $z:=\lceil \gamma|x| \rfloor \bmod q$.

\subsection{Keyed-Homomorphic Encryption}
To construct encrypted control systems, this study uses KHE with a multiplicative homomorphism \cite{Emu13}, which is currently the most efficient KHE scheme. 
The encryption scheme denoted as $\E=(\Gen,\Enc,\Dec,\Eval)$ consists of the following four algorithms.

\subsubsection{$\Gen$}$\kappa\mapsto (\pk, \skd, \skh)$. 
The $\Gen$ algorithm takes security parameter $\kappa$ and 
the key length $\ell$ regarding the $\ell$-bit prime number $p$ and outputs public, private, and homomorphic operation keys, denoted as $\pk$, $\skd$, and $\skh$, respectively:
$\pk=\left(g_{0},g_{1},s,\widehat{s}, \widetilde{s}_{0},\widetilde{s}_{1}\right)$, 
$\skd=\left(k_{0},k_{1},\widehat{k}_{0},\widehat{k}_{1},\widetilde{k}_{0,0},\widetilde{k}_{0,1},\widetilde{k}_{1,0},\widetilde{k}_{1,1}\right)$, and 
$\skh=\left(\widetilde{k}_{0,0},\widetilde{k}_{0,1},\widetilde{k}_{1,0},\widetilde{k}_{1,1}\right)$,
where $g_0$ and $g_1$ are randomly chosen from $\G$; 
$s:= g_{0}^{k_{0}}g_{1}^{k_{1}}\bmod p$;
$\widehat{s}:= g_{0}^{\widehat{k}_{0}}g_{1}^{\widehat{k}_{1}}\bmod p$;
$\widetilde{s}_{0}:= g_{0}^{\widetilde{k}_{0,0}}g_{1}^{\widetilde{k}_{0,1}}\bmod p$; $\widetilde{s}_{1}:= g_{0}^{\widetilde{k}_{1,0}}g_{1}^{\widetilde{k}_{1,1}}\bmod p$; 
$k_{0}$, $k_{1}$, $\widehat{k}_{0}$, $\widehat{k}_{1}$, $\widetilde{k}_{0,0}$,  $\widetilde{k}_{0,1}$, $\widetilde{k}_{1,0}$, and $\widetilde{k}_{1,1}$ are randomly chosen from $\Z_q$, where $p=2q+1$.

\subsubsection{$\Enc$}$(\pk, m\in\M)\mapsto c=(x_0,x_1,\epsilon,\widehat{\pi},\eta)\in\mathcal{C}$.
The $\Enc$ algorithm takes a public key $\pk$ and a plaintext $m$ and outputs a ciphertext $c$.
The components of $c$ are as follows:
$x_0:= g_0^\omega\bmod p$;
$x_1:= g_1^\omega\bmod p$; 
$\epsilon:= m\pi \bmod p$; 
$\widehat{\pi}:= \widehat{s}^{\omega}\bmod p$, where $\pi:= s^\omega\bmod p$ and $\omega$ is chosen randomly from $\mathbb{Z}_q$;
$\eta:= f_{h k}\left(\left(\widetilde{s}_{0} \cdot \widetilde{s}_{1}^{\delta}\right)^{\omega}\bmod p\right)$ with $\delta:= \Gamma_{h k}\left(x_{0}, x_{1}, \epsilon, \widehat{\pi}\right)$, 
where $\Gamma_{h k}$ is a target collision-resistance hash family and $f_{h k}$ is a smooth function. 
SHA-256 is used for both $\Gamma_{h k}$ and $f_{h k}$.

\subsubsection{$\Dec$}$(\skd,c\in\C)\mapsto m\in{\M}\cup\{\perp\}$. 
The $\Dec$ algorithm takes the private key $\skd$ and a ciphertext $c=(x_0,x_1,\epsilon,\widehat{\pi},\eta)$ and outputs a plaintext $m$ or an error symbol $\perp$.
First, compute: $\widehat{\pi}^{\prime}:= x_{0}^{\widehat{k}_{0}} x_{1}^{\widehat{k}_{1}}\bmod p$,
$\delta:= \Gamma_{h k}\left(x_{0}, x_{1}, \epsilon, \widehat{\pi}\right)$, and
$\eta^{\prime}:= f_{h k}\left(x_{0}^{\widetilde{k}_{0,0}+\delta \widetilde{k}_{1,0}} x_{1}^{\widetilde{k}_{0,1}+\delta \widetilde{k}_{1,1}}\bmod p\right)$, 
where $f_{h k}$ is a smooth function.
If either $\widehat{\pi} \neq \widehat{\pi}^{\prime}$ or $\eta \neq \eta^{\prime}$, then return an error symbol $\perp$; otherwise, return $m = \epsilon / \pi\bmod p$, 
where $\pi:=x_0^{k_0}x_1^{k_1} \bmod p$. 

\subsubsection{$\Eval$}$(\skh,c_1,c_2\in\C)\mapsto c\in\C\cup\{\perp\}$.
The $\Eval$ algorithm takes a homomorphic operation key and two ciphertexts $c_i$ $\forall i\in\{1,2\}$ and outputs a ciphertext $(x_0,x_1,\epsilon,\widehat{\pi},\eta)$ or an error symbol $\perp$. 
The components of the output $c$ are computed as follows: 
$x_{0}:= x_{1,0}x_{2,0}g_{0}^{\omega}\bmod p$, 
$x_{1}:= x_{1,1}x_{2,1}g_{1}^{\omega}\bmod p$,
$\epsilon:= \epsilon_{1}\epsilon_{2}s^{\omega}\bmod p$, 
$\widehat{\pi}:= \widehat{\pi}_{1}\widehat{\pi}_{2} \widehat{s}^{\omega}\bmod p$, and
$\eta = f_{h k}\left(x_{0}^{\widetilde{k}_{0,0}+\delta \widetilde{k}_{1,0}} x_{1}^{\widetilde{k}_{0,1}+\delta \widetilde{k}_{1,1}}\bmod p\right)$,
where 
$c_i:=\left(x_{i,0}, x_{i,1}, \epsilon_{i}, \widehat{\pi}_{i}, \eta_{i}\right)$; 
$\delta:= \Gamma_{h k}\left(x_{0}, x_{1}, \epsilon, \widehat{\pi}\right)$;
$\delta_i:= \Gamma_{h k}\left(x_{i,0}, x_{i,1}, \epsilon_i, \widehat{\pi}_i\right)$;
$\omega$ is randomly chosen from $\mathbb{Z}_q$;
$\eta_{i}^{\prime}:= f_{h k}\left(x_{i,0}^{\widetilde{k}_{0,0}+\delta_{i} \widetilde{k}_{1,0}} x_{i,1}^{\widetilde{k}_{0,1}+\delta_{i} \widetilde{k}_{1,1}}\bmod p\right)$.
If either $\eta_{1} \neq \eta_{1}^{\prime}$ or $\eta_{2} \neq \eta_{2}^{\prime}$, then return $\perp$; otherwise, return $c$.

The KHE scheme is known to satisfy the following two conditions:
i) For all $m\in\M$ and $c\in\C_{\pk,m}$, it holds that $\Dec(\skd,c)=m$;
ii) For all $m_1,m_2\in\M$, $c_1\in\C_{\pk,m_1}$, and $c_2\in\C_{\pk,m_2}$, it holds that $\Eval(\skh,c_1,c_2)\in\C_{\pk,m_1\cdot m_2}$,
where $\C_{\pk,m}$ denotes the set of all ciphertexts of $m\in\M$ under the public key $\pk$.
For simplicity, the arguments $\pk$, $\skd$, and $\skh$ will be omitted henceforth.

\section{Encrypting DOB-based PID Controller}\label{sec:prob}
This section formulates the encrypted DOB-based PID control system.

\begin{figure}[!t]
\centering
\includegraphics[width=0.95\columnwidth]{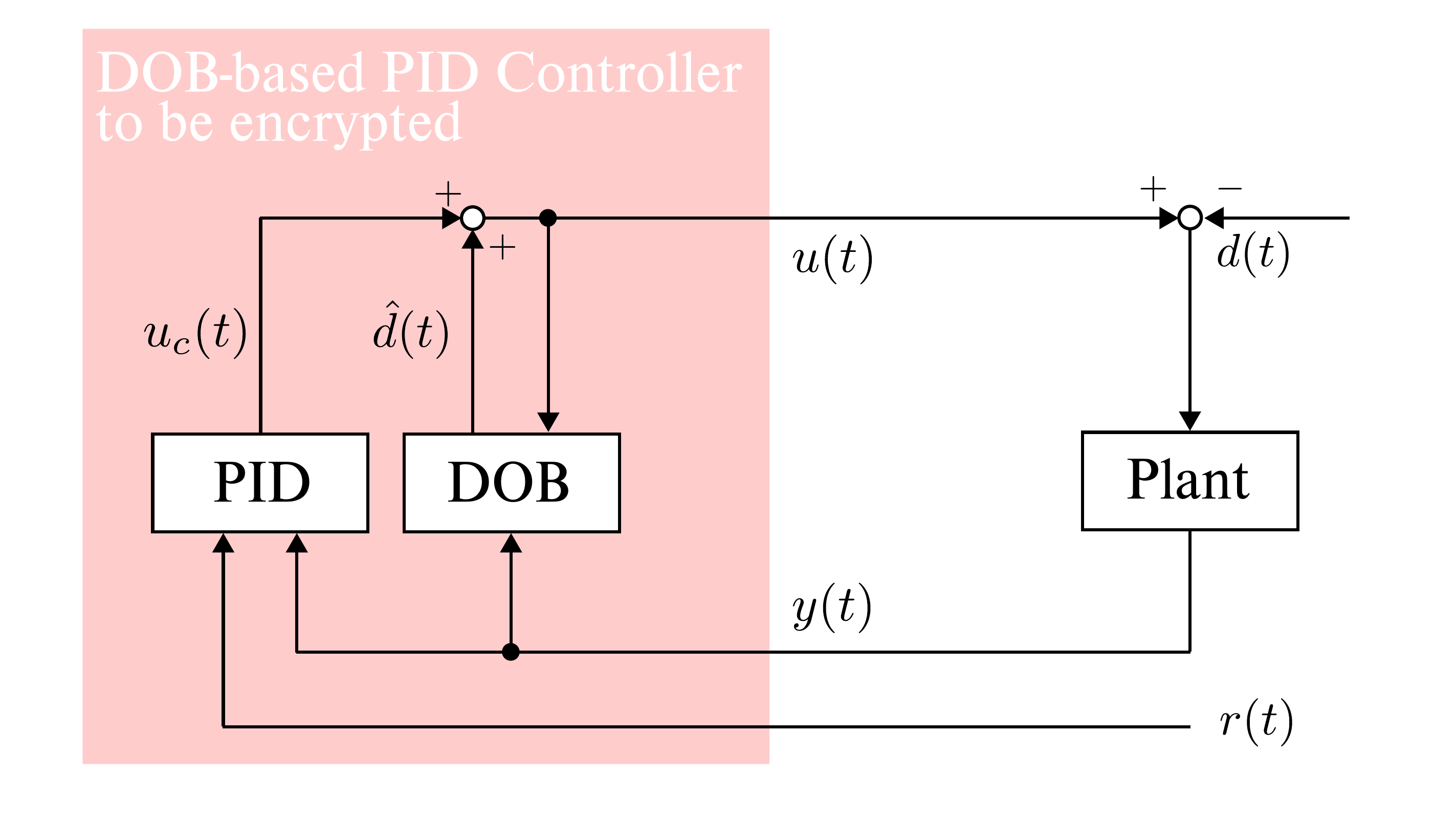}
\caption{Block diagram of the DOB-based PID control system}
\label{fig:block}
\end{figure}

\subsection{DOB-based PID controller}
This study considers the linear plant in the discrete-time state-space representation:
\begin{subequations}
\begin{align} 
x_{p}(t+1)&=A_{p}x_p(t)+B_{p}u(t)-B_{p}d(t),\\
y(t)&=C_{p}x_p(t), 
\end{align}\label{pmdl}
\end{subequations}
\!\!where 
$t\in\Z^+_0$ is the step, 
$x_p\in\R^2$ is the state, 
$u\in\R$ is the input, 
$y\in\R$ is the measurement (output), and 
$d\in\R$ is the exogenous disturbance, which is assumed to be unknown but constant over the steps.
Moreover, the pairs $(A_p,B_p)$ and $(C_p,A_p)$ are controllable and observable, respectively.

This study considers the situation where for the plant \eqref{pmdl}, the following DOB-based PID controller is designed, as illustrated in Fig.~\ref{fig:block}, to achieve the tracking control for a given setpoint reference $r\in\R$, 
\begin{subequations}
\begin{align}
x(t+1)&=Ax(t)+Bv(t),\\
u(t)&=Cx(t)+Dv(t),
\end{align}\label{ctrl}
\end{subequations}
\!\!where $x\in\R^5$ and $v\in\R^2$ are the controller’s state and input, respectively.
The definitions of the controller’s state and coefficients are derived below.

Firstly, the PID controller from a feedback error $e:=r-y\in\R$ to $u_c\in\R$:
\begin{align*}
u_c(t)=K_pe(t)+K_iw(t)+K_d\frac{e(t)-e(t-1)}{T_s},
\end{align*}
can form into the discrete-time state-space representation:
\begin{subequations}
\begin{align}
x_c(t+1)&=A_cx_c(t)+B_cv(t),\label{pid_a}\\
u_c(t)&=C_cx_c(t)+D_cv(t), \label{pid_b}
\end{align}\label{pid}
\end{subequations}
\!\!where 
$x_c(t):=\begin{bmatrix}e(t-1) & w(t-1)\end{bmatrix}^\top\in\R^2$ and 
$v(t):=\begin{bmatrix}r(t) & y(t)\end{bmatrix}^\top\in\R^2$ are the PID-controller’s state and input, respectively; 
$w\in\R$ is the accumulated error, i.e., $w(t)=T_s\sum_{\tau=0}^{t-1}e(\tau)=w(t-1)+T_se(t-1),\,\forall t\in\Z_0^+$, where $e(-1)$ and $w(-1)$ are set to zero;
$K_p$, $K_i$, and $K_d$ are the PID gains; 
$T_s>0$ is the sampling period;
The coefficients in~\eqref{pid} are as follows:
\begin{align*}
    A_c&=
    \begin{bmatrix}
        0 & 0 \\ 0 & 1
    \end{bmatrix},\ 
    B_c=
    \begin{bmatrix}
        1 & -1 \\ T_s & -T_s
    \end{bmatrix},\ 
    C_c=
    \begin{bmatrix}
        -\frac{K_d}{T_s} & K_i
    \end{bmatrix}, \\
    D_c&=
    \begin{bmatrix}
        K_p + K_{i}T_{s}+\frac{K_d}{T_s} & -(K_p + K_{i}T_{s}+\frac{K_d}{T_s})
    \end{bmatrix}.
\end{align*}

Subsequently, the DOB is given as follows:
\begin{subequations}
\begin{align}
x_d(t+1)
&=
\begin{bmatrix}
A_p&-B_p\\ \textbf{0}&1
\end{bmatrix}
x_d(t)+
\begin{bmatrix}
B_p\\0
\end{bmatrix}
u(t)+ \nonumber\\& \quad\,
\begin{bmatrix}
L_x\\L_d
\end{bmatrix}
(y(t)-\hat{y}(t)), \label{dob_a}\\
\begin{bmatrix}\hat{y}(t)\\\hat{d}(t)\end{bmatrix}&=
\begin{bmatrix}C_p & 0\\
\textbf{0}&1\end{bmatrix}x_d(t), \label{dob_b}
\end{align}\label{dob}
\end{subequations}
\!\!where $x_d:=\begin{bmatrix}\hat{x}_p^\top & \hat{d}\end{bmatrix}^\top\in\R^3$ is the DOB's state consisting of the estimate $\hat{x}_p$ of the plant state $x_p$ and the estimated disturbance $\hat{d}$; 
$\hat{y}\in\R$ is the estimated plant's output; 
$L_x\in\R^{2}$ and $L_d\in\R$ are the observer gains.
Furthermore, using $u_c$ in \eqref{pid_b} and $\hat{d}$ in \eqref{dob_b}, the control input $u$, which is the output of the DOB-based PID controller, is given as follows:
\begin{align}
u(t)&=u_c(t)+\hat{d}(t),\nonumber\\[.6ex]
&=C_cx_c(t)+D_cv(t)+\begin{bmatrix}\textbf{0}&1\end{bmatrix}x_d(t).\label{cctrl}
\end{align}
The state equation of the DOB-based PID controller consists of \eqref{pid_a} and \eqref{dob_a} after eliminating $u$ and $\hat{y}$ from~\eqref{dob_a}.

Consequently, defining the state in \eqref{ctrl} as $x:=\begin{bmatrix}x_c^\top&x_d^\top\end{bmatrix}^\top$, the DOB-based PID controller is described as follows:
\begin{subequations}
\begin{align}
x(t+1)&=
\begin{bmatrix}
A_c & \textbf{0}\\ B_d & A_d
\end{bmatrix}
x(t)+
\begin{bmatrix}
B_c\\ C_d
\end{bmatrix}v(t),\\[.6ex]
u(t)&=
\begin{bmatrix}
C_c & \begin{matrix}\textbf{0}&1\end{matrix}
\end{bmatrix}
x(t)+D_c v(t),
\end{align}\label{ctrl2}
\end{subequations}
\!\!where $A_d\in\R^{3\times3}$, $B_d\in\R^{3\times2}$, and $C_d\in\R^{3\times2}$ are as follows:
\begin{align*}
     A_d &=
    \begin{bmatrix}
        A_p -L_xC_p & \textbf{0} \\ -L_dC_p & 1
    \end{bmatrix},\ 
    B_d=
    \begin{bmatrix}
        B_pC_c \\ \textbf{0}
    \end{bmatrix},\\
    C_d &=
    \begin{bmatrix}
        B_p(K_p+K_iT_s+\frac{K_d}{T_s}) & -B_p(K_p+K_iT_s+\frac{K_d}{T_s}) \\ 0 & L_d
    \end{bmatrix}.
\end{align*}
Since \eqref{ctrl} and \eqref{ctrl2} are identical, the coefficients $(A,B,C,D)$ result in
\begin{align*}
A=\begin{bmatrix}
A_c & \textbf{0}\\ B_d & A_d
\end{bmatrix},\, 
B=\begin{bmatrix}
B_c\\ C_d
\end{bmatrix},\,
C=\begin{bmatrix}
C_c & \begin{matrix}\textbf{0}&1\end{matrix}
\end{bmatrix},\,
D=D_c.
\end{align*}


\subsection{Controller Encryption}
The used KHE scheme is the type of multiplicatively homomorphic encryption, so the controller encryption method, proposed in~\cite{Kogiso15}, can be applied to secure implementation of the controller \eqref{ctrl}.
It is rewritten as follows:
\begin{align}
\psi(t)=\Phi\xi(t)=:f(\Phi,\xi(t)),\label{ctrlr}
\end{align}
where $\psi\in\R^6$, $\Phi\in\R^{6\times 7}$, and $\xi\in\R^7$ are as follows,
\begin{align*}
\psi(t)&:=\begin{bmatrix} x(t+1) \\ u(t)\end{bmatrix}
=\begin{bmatrix}e(t) \\ w(t) \\ \hat{x}_p(t+1) \\ \hat{d}(t+1) \\ u(t)\end{bmatrix},\ 
\Phi:=\begin{bmatrix}A & B\\ C & D\end{bmatrix},\\[.6ex]
&\,\xi(t):=\begin{bmatrix}x(t) \\ v(t)\end{bmatrix}
=\begin{bmatrix}e(t-1) \\ w(t-1) \\ \hat{x}_p(t) \\ \hat{d}(t) \\ r(t) \\ y(t)\end{bmatrix}.
\end{align*}

\begin{figure}[!t]
\centering
\includegraphics[width=0.95\columnwidth]{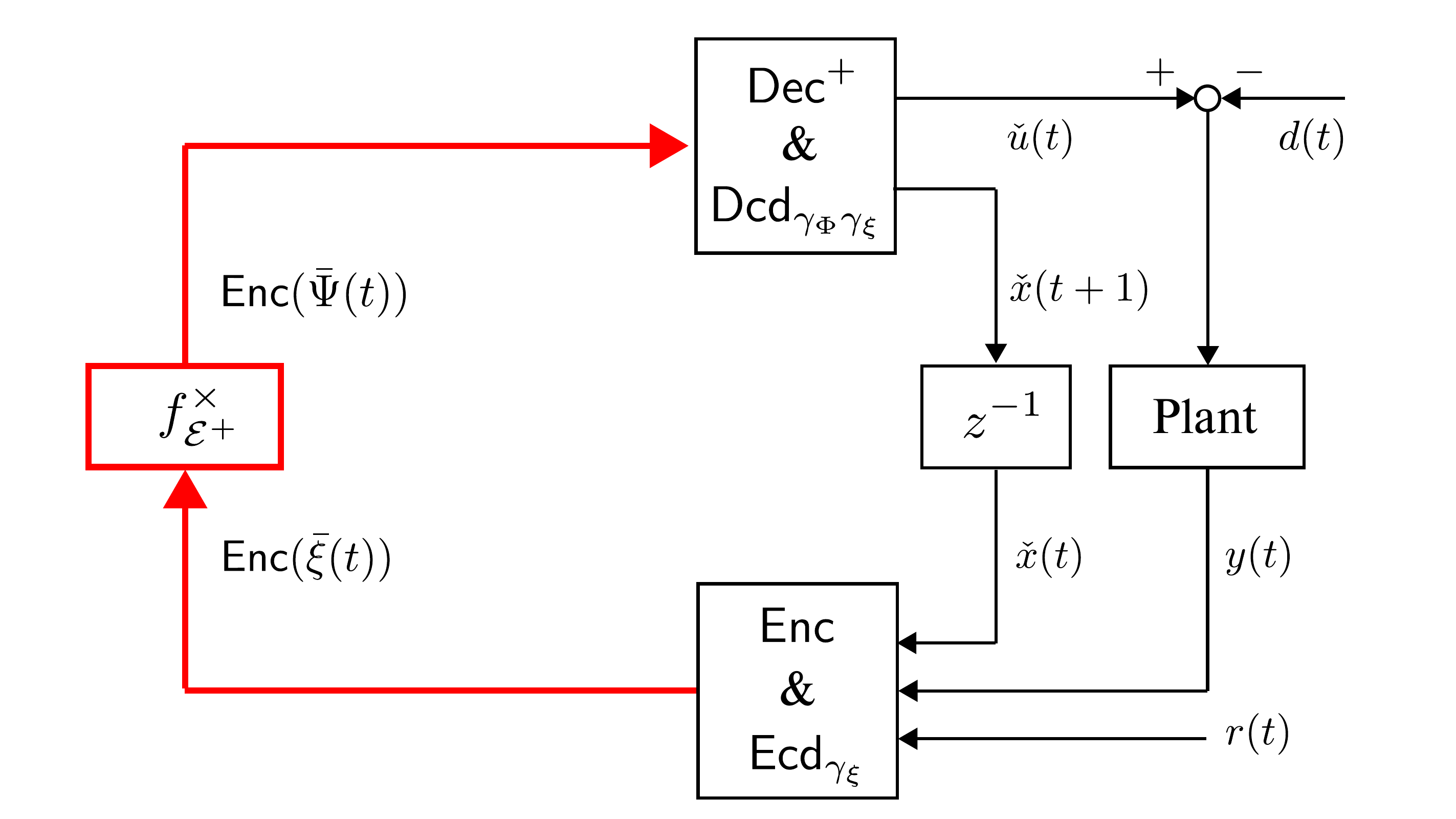}
\caption{Block diagram of the proposed encrypted DOB-based PID control system.}
\label{fig:block_ecs}
\end{figure}

For the DOB-based PID controller \eqref{ctrlr}, since 
$f$ is a composition product of multiplication $f^{\times}$ and addition $f^{+}$, the decryption algorithm is modified to yield $\Dec^{+}=f^{+}\circ\Dec$ \cite{Kogiso15}.
The modified homomorphic encryption scheme $\E^{+}=(\Gen,\Enc,\Dec^{+}, \Eval)$ enables to construct the encrypted controller $f_{\E^{+}}^{\times}$ as follows:
\begin{align}
    f_{\E^{+}}^{\times}:(\Enc({\bar{\Phi}}),\Enc(\bar{\xi}(t))\mapsto \Enc(\bar{\Psi}(t)),\label{eq:ectrl}
\end{align}
where
$\bar{\Phi}=\Ecd_{\gamma_{\Phi}}(\Phi)$, $\bar{\xi}=\Ecd_{\gamma_{\xi}}(\xi)$, $\bar{\Psi}=\Ecd_{\gamma_{\Phi}\gamma_{\xi}}(f^{\times}(\Phi,\xi))$, $\gamma_{\Phi}$ and $\gamma_{\xi}$ are quantization gains regarding $\Phi$ and $\xi$, respectively, and $\Enc(\bar{\Psi}(t))$ is calculated as follows:
\begin{align}
  \Enc(\bar{\Psi}_{ij}^{\theta}(t)) &= \Eval(\Enc({\bar{\Phi}_{ij}^\theta},\Enc({\bar{\xi}_{j}^{\theta}}(t))), \nonumber\\
  &\hspace*{5ex} \forall \theta\in\{1,2\}, \forall i\in\mathbb{Z}^+_{7}, \forall j\in\mathbb{Z}^+_{8}. \label{eq:eval}
\end{align}
The function \eqref{eq:ectrl} is a ciphertext version of \eqref{ctrlr}, which is realized in \eqref{eq:eval}.
Therefore, as shown in Fig.~\ref{fig:block_ecs}, function $f$ running in the controller is replaced by $f_{\E^{+}}^{\times}$, and the controller output $\Enc({\bar{\Psi}}(t))$ is decrypted and decoded at the plant side to extract control input $u$ via a decoded (quantized) signal $\check{\psi}$:
\begin{align*}
\check{\psi}=\begin{bmatrix}\check{x}(t+1) \\ \check{u}(t)\end{bmatrix}:=\Dcd_{\gamma_\Phi \gamma_\xi}(\Dec^{+}(\Enc({\bar{\Psi}}(t)))).
\end{align*}
Additionally, $u-\check{u}$ means the quantization error in $u$.

The next section will discuss the effectiveness of the encrypted DOB-based PID control presented as a cybersecurity measure, using an industrial linear stage.

\begin{remark}
This study assumes that transmission delays between the plant and controller sides are sufficiently shorter than a sampling period to simplify the discussion. 
This implies that communication links can be used without their quality affecting control performance.
\end{remark}

\section{Experimental Validation} \label{sec:ev}
This section validates the encrypted DOB-based PID control of an industrial linear stage, compared with the encrypted PID control addressed in~\cite{Miy23}.

\subsection{Positioning Control System for Linear Stage}
\begin{figure}[!t]
\centering
\includegraphics[width=0.95\columnwidth]{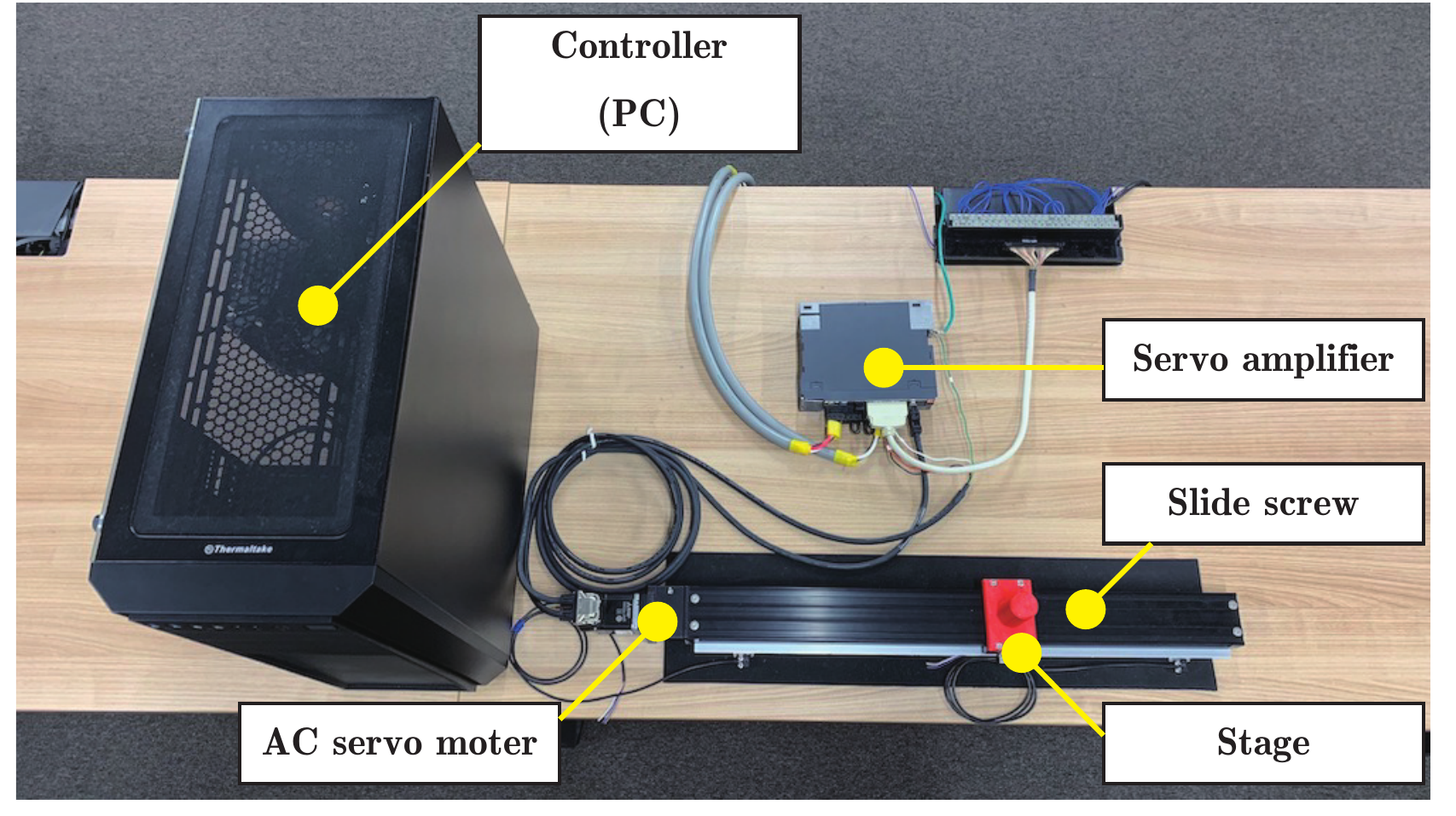}
\caption{Whole view of positioning control system for industrial linear stage~\cite{Miy23}.}
\label{exp:cs}
\end{figure}

\begin{figure}[!t]
  \begin{minipage}[t]{0.49\linewidth}
    \centering
    \includegraphics[keepaspectratio, scale=0.49]{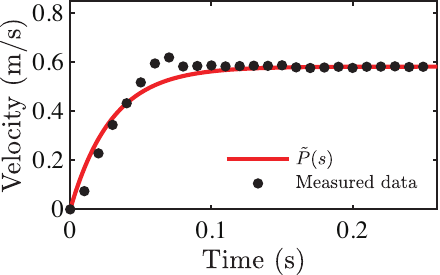}
    \subcaption{Velocity output $v(t)$}\label{fig:stepv}
  \end{minipage}
  \begin{minipage}[t]{0.49\linewidth}
    \centering
    \includegraphics[keepaspectratio, scale=0.49]{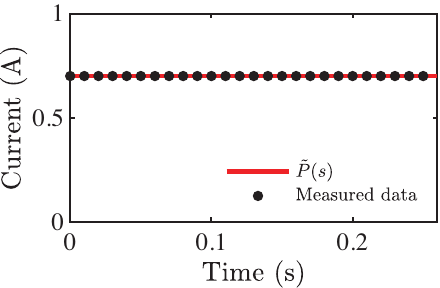}
    \subcaption{Input current $u(t) $}\label{fig:stepu}
  \end{minipage}
\caption{Model identification by step-response experiment.}
\label{fig:step}
\end{figure}

This study considers the DOB-based positioning control system for the linear stage, as shown in Fig.~\ref{exp:cs}.
The stage and the computer setup are the same as those used in~\cite{Miy23}, with their specifications summarized in TABLE~\ref{tab:spec}.
The input to the stage is current (A), and its output is position measured by a linear encoder. 
The stage model is expressed as $P(s)=s^{-1}\tilde{P}(s)$, where $\tilde{P}(s)$ captures the dynamics from the current input to velocity.
Based on a step-response experiment, the dynamics were identified as: 
\begin{align}
\tilde{P}(s)=\frac{28.288}{s+34}, \nonumber
\end{align}
where the step input was $u(t)=0.7$.
The results of the model identification are shown in Fig.~\ref{fig:step}.
Figs.~\ref{fig:step}(a) and (b) illustrate the time responses of the stage velocity and control input, respectively.
In these figures, black dots represent the measured velocity, and the red line corresponds to the output of $\tilde{P}(s)$.
Defining the plant state as $x_p\in\Re^2$ and discretizing the dynamics using a zero-order hold with a sampling period $T_s=10$~ms, the discrete-time state-space representation of $P(s)$ is realized with the following system matrices:
\begin{align*}
    A=
    \begin{bmatrix}
        1 & 0.0085 \\ 0 & 0.7118
    \end{bmatrix}, ~
    B=
    \begin{bmatrix}
        0.0013 \\ 0.2398
    \end{bmatrix}, ~
    C=
    \begin{bmatrix}
        1 & 0
    \end{bmatrix},
\end{align*}
where the state is defined as $x_p:=\begin{bmatrix}x_{p,1} \, x_{p,2}\end{bmatrix}^\top$, with $x_{p,1}$ and $x_{p,2}$ representing the stage position (m) and velocity (m/s), repsectivey.

The observer gains were designed as follows: 
\begin{align}
L_{x}=\begin{bmatrix}2.7118 & 199.2800\end{bmatrix}^{\top}, \quad L_{d}=-380.4456, \nonumber
\end{align}
ensuring that the eigenvalues of the DOB are smaller than those of the plant.
The position reference was given as: 
\begin{align}
r(t)=
\begin{cases}
\,0 & \textrm{if}\ \ 0 \leq T_st < 2.0,\\
\,0.05 & \textrm{if}\ \ 2.0 \leq T_st < 4.0,\\
\,0.10 & \textrm{if}\ \ 4.0 \leq T_st < 6.0,\\
\,0.05 & \textrm{if}\ \ 6.0 \leq T_st < 8.0,\\
\,0 & \textrm{if}\ \ 8.0 \leq T_st.
\end{cases}\label{ref}
\end{align}
Based on the control objective of tracking the reference, the PID controller parameters were tuned through trial and error and set as follows: $K_p=12$, $K_i=0.25$, and $K_d=0.030$.
In this case, $\Phi$ is given as follows:
\vspace*{-1ex}
\begin{align*}
&\Phi=\\
&{\scriptsize
\begin{bmatrix}       0&0&0&0&0&1&-1\\0&1&0&0&0&0.01&-0.01\\-0.0039&0.0003&-1.7118&0.0085&0&0.0195&2.6923\\-0.7194&0.0600&-199.2800&0.7118&0&3.5976&195.6824\\0&0&380.4456&0&1&0&-380.4456\\-3&0.25&0&0&1&15.0025&-15.0025
    \end{bmatrix}}
\end{align*}

\begin{figure}[!t]
  \begin{minipage}[t]{0.49\linewidth}
    \centering
    \includegraphics[keepaspectratio, scale=0.49]{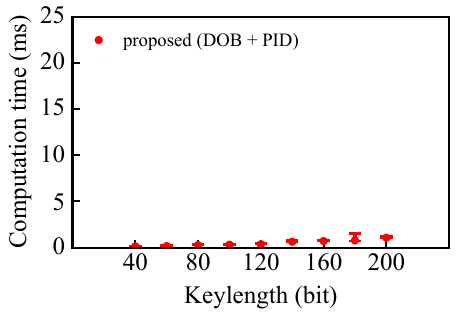}
    \subcaption{ $\mathsf{Enc}$}\label{fig:tenc}
  \end{minipage}
  \begin{minipage}[t]{0.49\linewidth}
    \centering
    \includegraphics[keepaspectratio, scale=0.49]{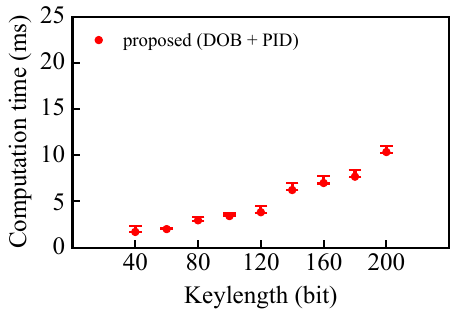}
    \subcaption{ $\mathsf{Eval}$}\label{fig:teval}
  \end{minipage}
  \begin{minipage}[t]{0.49\linewidth}
    \centering
    \includegraphics[keepaspectratio, scale=0.49]{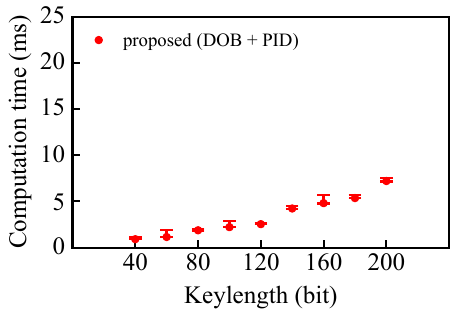}
    \subcaption{ $\mathsf{Dec}^{+}$}\label{fig:tdec}
  \end{minipage}
  \begin{minipage}[t]{0.49\linewidth}
    \centering
    \includegraphics[keepaspectratio, scale=0.49]{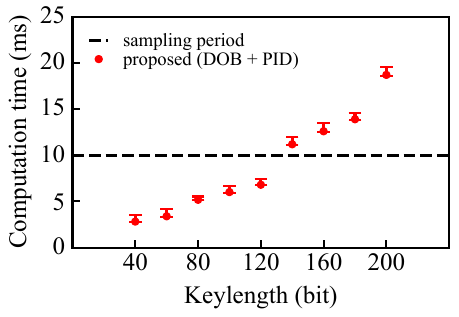}
    \subcaption{Total time}\label{fig:ttotal}
  \end{minipage}
\caption{Computation time of each process over the key length.}
\label{fig:time}
\end{figure}

The key length was set to 120 bits for encrypting the controller.
This value was determined by evaluating the computation time of the encrypted control processes, including $\Enc$, $\Eval$, $\Dec^{+}$ across different key lengths. 
The average computation times, measured over 1,000 trials for key lengths ranging from 40 to 200 bits in increments of 20 bits, are presented in Fig.~\ref{fig:time}.
Figs.~\ref{fig:time}(a), (b), and (c) show the computation times for $\Enc$, $\Eval$, and $\Dec^{+}$, respectively, and Fig.~\ref{fig:time}(d) shows the total computation time. 
The red dots represent the proposed control systems.
The key length was set to 120 bits based on the observation that the average computation time at this length is 6.806 ms, with a variation of 0.685 ms between the maximum and minimum values, which is less than $T_s$.
This configuration ensures real-time performance.
Furthermore, for preventing overflow, quantization gains were set to $\gamma_{\Phi}=10^{15}$ and $\gamma_{\xi}=10^{16}$, determined through trial and error based on Theorem 3.3 in~\cite{Miy23}.

\begin{remark}
To ensure the security of the KHE scheme used in this study, the key length needs to be 2048 bits, as recommended by the NIST document~\cite{Bar20}.
However, such a long key can impact the efficiency of the control system, making alternative approaches, such as key updates, a potential solution.
Additionally, the current static key setting allows replay attacks despite the detection mechanism.
Enhancing security without compromising efficiency, for instance, through key updates, remains an important direction for future work.
\end{remark}

\subsection{Experimental Control Results and Evaluations}
This section addresses two methods: One is the proposed encrypted DOB-based PID control method, and the other is the conventional encrypted PID control method~\cite{Miy23}.
The following time-averaged $\ell_1$-norm index is introduced to measure the tracking performance between 2~s and 10~s:
\begin{align}
    \rho(e)&:=\frac{1}{10^3-200}\sum_{t=200}^{10^3-1} |r(t)-y(t)|, \nonumber
\end{align}
where $e:=r-y$ is the feedback (tracking) error, and $|\cdot|$ denotes the absolute value operator.

\begin{figure}[!t]
  \begin{minipage}[t]{0.49\linewidth}
    \centering
    \includegraphics[keepaspectratio, scale=0.49]{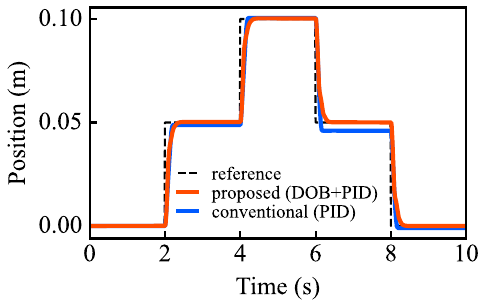}
    \subcaption{Output $y(t)$}\label{fig:y_dob}
  \end{minipage}
  \begin{minipage}[t]{0.49\linewidth}
    \centering
    \includegraphics[keepaspectratio, scale=0.49]{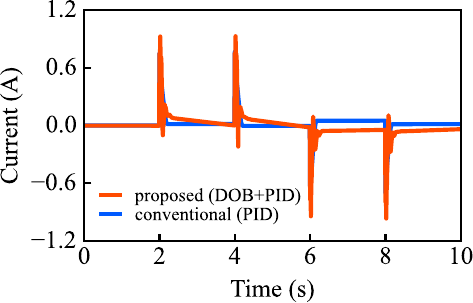}
    \subcaption{Input $u(t)$}\label{fig:u_dob}
  \end{minipage}
  \begin{minipage}[t]{0.49\linewidth}
    \centering
    \includegraphics[keepaspectratio, scale=0.49]{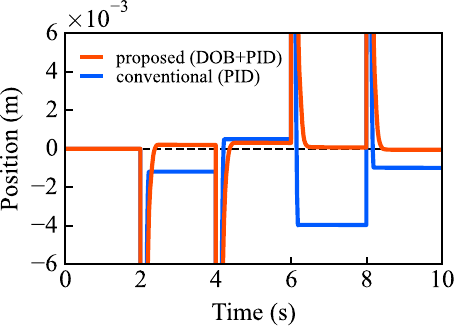}
    \subcaption{Tracking error $e(t)$}\label{fig:ed_dob}
  \end{minipage}
    \begin{minipage}[t]{0.49\linewidth}
    \centering
    \includegraphics[keepaspectratio, scale=0.49]{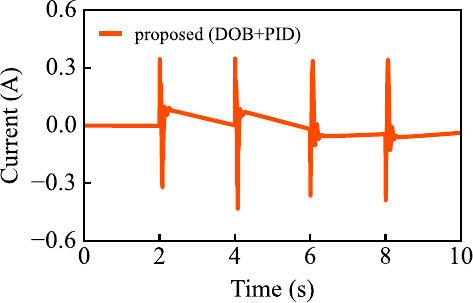}
    \subcaption{Estimated disturbance $\hat{d}(t)$}\label{fig:qe_dob}
  \end{minipage}
  \begin{minipage}[t]{0.49\linewidth}
    \centering
    \includegraphics[keepaspectratio, scale=0.49]{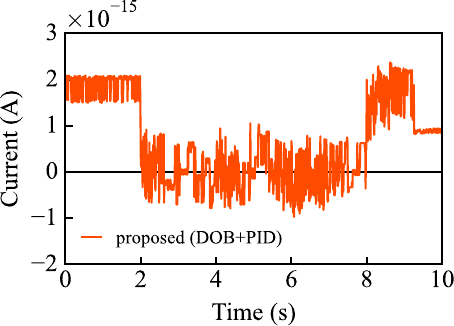}
    \subcaption{Quantization error $u(t)-\check{u}(t)$ of the proposed  controller}\label{fig:te_ecs}
  \end{minipage}
  \begin{minipage}[t]{0.49\linewidth}
    \centering
    \includegraphics[keepaspectratio, scale=0.49]{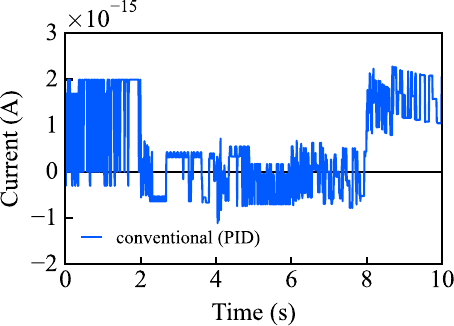}
    \subcaption{Quantization error $u(t)-\check{u}(t)$ of the conventional controller \cite{Miy23}}\label{fig:te2_ecs}
  \end{minipage}
   \begin{minipage}[t]{0.49\linewidth}
    \centering
    \includegraphics[keepaspectratio, scale=0.49]{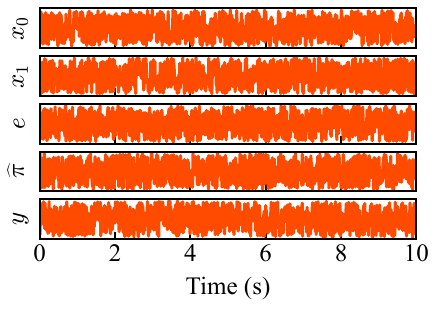}
    \subcaption{$\Enc(\bar{y}^1(t))$}\label{fig:ency1_dob}
  \end{minipage}
  \begin{minipage}[t]{0.49\linewidth}
    \centering
    \includegraphics[keepaspectratio, scale=0.49]{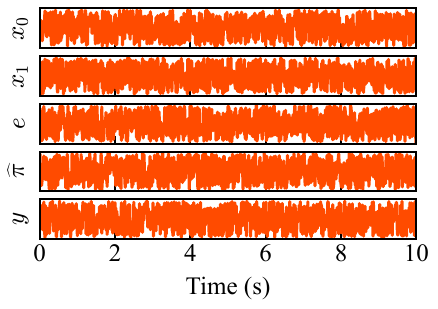}
    \subcaption{$\Enc(\bar{y}^2(t))$}\label{fig:ency2_dob}
  \end{minipage}
\caption{Experimental results of the proposed and conventional encrypted controls.}
\label{fig:ecs}
\vspace*{-2ex}
\end{figure}

The positioning control results of the proposed control system and the conventional control system are shown in Fig.~\ref{fig:ecs}.
Figs.~\ref{fig:ecs}(a) and (b) show the time responses of the measured stage position and the control input, respectively, and Fig.~\ref{fig:ecs}(c) shows the tracking error.
In these figures, the red and blue lines represent the proposed control system and the conventional control system, respectively, and the black broken line represents the reference~\eqref{ref}.
Fig.~\ref{fig:ecs}(d) shows the time response of the disturbance estimated by the DOB.
Figs.~\ref{fig:ecs}(e) and (f) show the time responses of the quantization error $u(t)-\check{u}(t)$ of the proposed control system and the conventional control system, respectively.
Figs.~\ref{fig:ecs}(g) and (h) show the time responses of the encrypted output $\Enc(\bar{\xi}_7^1(t))$ and $\Enc(\bar{\xi}_7^2(t))$, respectively, where $\xi_7$ is corresponding to the stage position $y$ in the proposed control system. 

In Fig.~\ref{fig:ecs}(c), The tracking performance index scores $\rho(e)$ for the proposed and the conventional control system are $2.669\times 10^{-3}$ and $3.737\times 10^{-3}$, respectively.
The results demonstrate that the proposed control system improves tracking performance by compensating for the disturbance through DOB.
Figs.~\ref{fig:ecs}(e) and (f) confirm that the quantization errors induced in both the proposed and the conventional control systems are negligible because it is in the $10^{-15}$ ampere range and too small to affect the stage. 
Furthermore, Figs.~\ref{fig:ecs}(g) and (h) validate that the stage position $y$ is effectively concealed by random numbers.
Therefore, these experimental control results confirm that the proposed control system achieves better tracking performance than the conventional control system.

\begin{table}[t!]
\centering
\caption{Experimental apparatus~\cite{Miy23}}
\label{tab:spec}
\begin{tabular}{ll}
\hline
\textbf{Servo amplifier}  & MITSUBISHI MR-J5-10A \\
Main circuit power supply & 1/3-phase 200-240 VAC 50/60 Hz                    \\
\textbf{AC servo motor}   & MITSUBISHI HK-KT13W  \\
Rated power               & 0.1 kW               \\
Rated torque              & 0.32 Nm             \\
Rated speed               & 3000 rpm           \\
Rated current             & 1.2 A                \\
Pulse per rotation        & 67108864 ppr                    \\ \hline
\textbf{Slide screw}      & MiSUMi LX3010CP-MX   \\
Length                    & 1250 mm              \\
Lead                      & 10 mm                \\ \hline
\textbf{PC}               &                      \\
CPU                       & Intel Core i7-10700K 3.80 GHz    \\
Memory                    & 64 GB    \\
OS                        & CentOS Linux 8       \\
Language                  & C++17      \\
DA/AD board               & Interface PEX-340216 (16-bit resolution)\\
Counter board             & Interface PEX-632104 (32-bit resolution)\\\hline
\end{tabular}
\end{table}

\section{Demonstration Of Attack Detection}\label{sec:ad}
This section demonstrates the effectiveness of malleability-based falsification attack detection through experimental attack tests.

\subsection{Attack Models}
In this study, we assume that the attackers have internal access to the controller and understand the structure of the ciphertext $c$, as well as the method to compute the plaintext $m$ from $c$, i.e., $m=\epsilon(\pi)^{-1}\bmod p$.
However, attackers do not possess the private or homomorphic operation keys, $\skd$ and $\skh$.
During a falsification attack, attackers overwrite the third component of the encrypted data $c$ by multiplying it with $\lambda\in\G$, as follows:
\begin{align*}
c^a=(x_{0},x_{1},\lambda\epsilon, \widehat{\pi}, \eta)\in\C.
\end{align*}
As a result, the falsified ciphertext $c^a$ leads to $\lambda\epsilon(\pi)^{-1}\bmod p=\lambda m$.
Even without knowledge of the private key, attackers can manipulate the third component of the ciphertext to scale the control parameter by an integer factor of $\lambda$.
However, if the $\Eval$ and $\Dec$ algorithms are correctly implemented, such manipulation will result in the output of an error symbol.

This study examines two falsification attack scenarios targeting control parameters:
i) Case 1: The 61th component of the encrypted system matrix $\Phi$ corresponding to the control parameters, denoted as $\Phi_{61}$, is multiplied by $\lambda=12$:
\begin{align*}
c_{\bar{\Phi}_{61}^{2}}(t)=
  \begin{cases}
   c_{\bar{\Phi}_{61}^{2}}^a(t),\ &\textrm{if}\ \ T_{s}t \in [5,  10),\\
   c_{\bar{\Phi}_{61}^{2}}(t), &\text{otherwise}.
  \end{cases}
\end{align*}
Because $\Phi_{61}^{2}=C_{11}$, multiplying $\Phi_{61}^{2}$ by 12 inceases $u$, leading to degraded control performance.
Additionally, because attackers can regulate $u$ using $\lambda$, the attack is difficult to detect using threshold-based detection methods, such as the one presented in \cite{baba18}, and
ii) Case 2: This scenario considers an attack that destabilizes the control system by modifying~$\Phi_{55}$.
Specifically, $\Phi_{55}$ is multiplied by $\lambda=2$:
\begin{align*}
c_{\bar{\Phi}_{55}^{2}}(t)=
  \begin{cases}
   c_{\bar{\Phi}_{55}^{2}}^a(t),\ &\textrm{if}\ \ T_{s}t \in [5,  10),\\
   c_{\bar{\Phi}_{55}^{2}}(t), &\text{otherwise}.
  \end{cases}
\end{align*}
As a result, the absolute values of the control system’s eigenvalues change from $\{$0.3571, 0.3571, 0.0083, 0.1768, 0.9618, 0.9618, 0.9998$\}$ to  
$\{$0.5476, 0.5476, 1.7145, 0.0051, 0.5570, 1.2074, 1.0001$\}$, 
which indicates that the attack destabilizes the system.

\begin{figure}[!t]
  \begin{minipage}[t]{0.49\linewidth}
    \centering
    \includegraphics[keepaspectratio, scale=0.49]{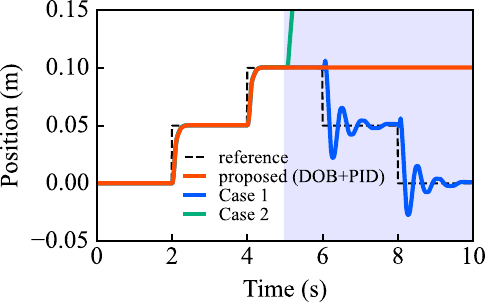}
    \subcaption{Output $y(t)$}\label{fig:y_fal}
  \end{minipage}
  \begin{minipage}[t]{0.49\linewidth}
    \centering
    \includegraphics[keepaspectratio, scale=0.49]{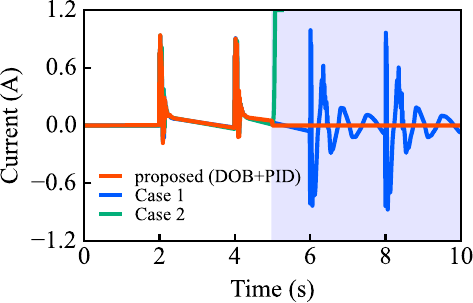}
    \subcaption{Input $u(t)$}\label{fig:u_fal}
  \end{minipage}
  \begin{minipage}[t]{0.49\linewidth}
    \centering
    \includegraphics[keepaspectratio, scale=0.49]{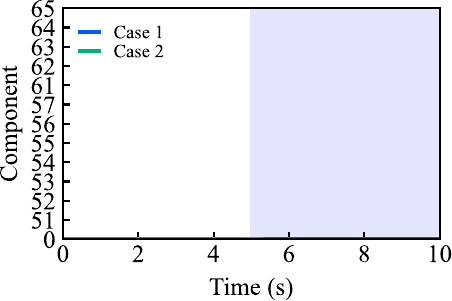}
    \subcaption{Error symbol for $\Enc(\bar{\Psi}^1)$}\label{fig:dt1}
  \end{minipage}
  \begin{minipage}[t]{0.49\linewidth}
    \centering
    \includegraphics[keepaspectratio, scale=0.49]{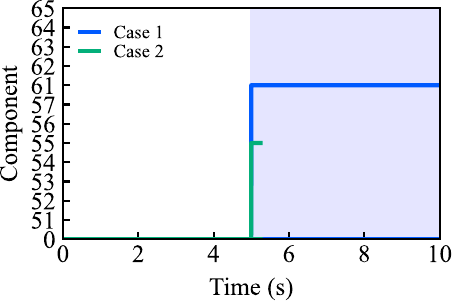}
    \subcaption{Error symbol for $\Enc(\bar{\Psi}^2)$}\label{fig:dt2}
  \end{minipage}
\caption{Results of falsification attack against the proposed encrypted control system}
\label{fig:fal}
\vspace*{-2ex}
\end{figure}

\subsection{Detection Results}
In the experiment, the control input is set to zero whenever an error symbol is detected in the components of $C_{\bar{\Psi}}$ because the stage can accelerate rapidly and potentially damage the experimental setup.
The control system and objectives remain identical to those in Section~\ref{sec:ev}.

The detection results of the proposed encrypted control system under falsification attacks are shown in Fig.~\ref{fig:fal}.
Figs.~\ref{fig:fal}(a) and (b) show the time responses of the measured stage position and the control input, respectively.
Figs.~\ref{fig:fal}(c) and (d) show the error symbols for 51th to 65th components of $C_{\bar{\Psi}^{1}}$ and $C_{\bar{\Psi}^{2}}$ in $\Eval$, respectively.
In these figures, the red, blue, and green lines correspond to: 
i) The proposed control system against the attack in both Case 1 and Case 2 (results are identical for both cases), 
ii) The response when the proposed control system ignores the detection in Case 1, and 
iii) The response when the proposed control system ignores the detection in Case 2, respectively.
The blue-shaded area indicates the duration of the falsification attack.
In Case 2, the stage was stopped at 5.32 s by activating the emergency stop button due to its rapid acceleration.

Figs.~\ref{fig:fal}(d) confirms that the indices of the attacked components, $C_{\bar{\Psi}_{61}^{2}}$ and $C_{\bar{\Psi}_{55}^{2}}$, are successfully detected by the proposed control system.
Consequently, the control input $u$ was set to zero during the attack period, maintaining the stage position, as shown by the red line of Fig.~\ref{fig:fal}(a).
These results demonstrate that the proposed encrypted control system can detect falsification attacks in real time and accurately identify the affected components.

\section{Conclusion}\label{sec:con}
This study proposed an encrypted DOB-based PID control system using a KHE scheme, aiming to achieve control performance and resistance to malleability-based attacks. 
The system was experimentally validated on an industrial linear stage through both tracking control tests and attack detection experiments. 
The results demonstrated that the proposed controller outperformed a conventional encrypted PID controller in terms of tracking accuracy while maintaining the confidentiality of communication signals and control parameters. 
Furthermore, the attack detection experiments confirmed that the system could successfully identify malleability-based falsifications and localize the compromised components in real time.

For future work, we plan to explore the integration of updatable KHE to further enhance both the security and computational efficiency of encrypted control systems. 
This extension would enable the detection of replay attacks and support longer key lengths while maintaining real-time feasibility. 
We also intend to investigate countermeasures against the leakage of homomorphic operation keys to ensure security even under partial exposure of cryptographic information. 
In addition, developing methods to estimate or detect the manipulation factor $\lambda$ would enhance system resilience by enabling real-time identification of unauthorized modifications. 
Finally, exploring (attribute-based) keyed fully homomorphic encryption schemes~\cite{Sat22,Emu24} represents a promising direction for advancing the design and capabilities of encrypted control systems.


 





\begin{IEEEbiography}[{\includegraphics[width=1in,height=1.25in,clip,keepaspectratio]{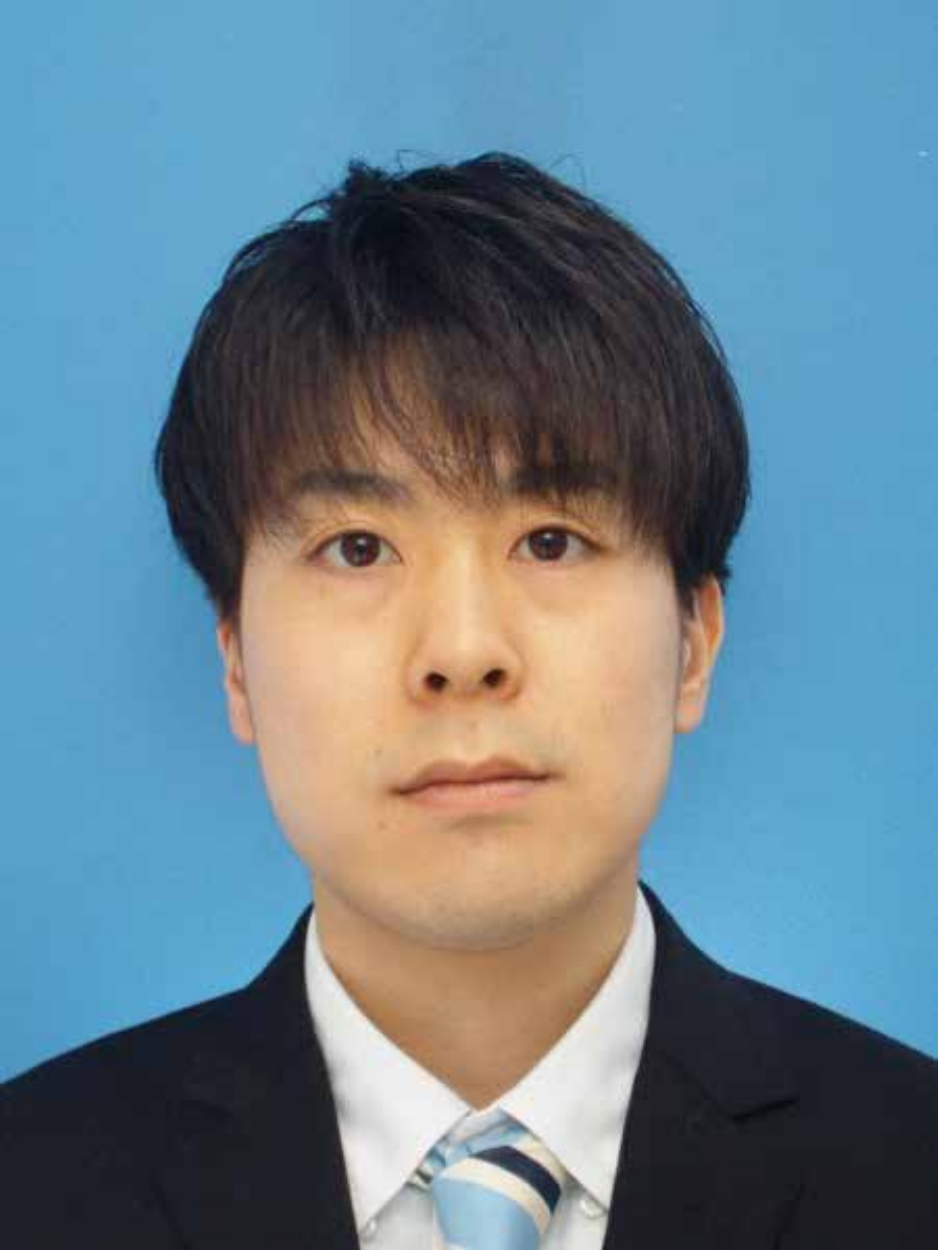}}]
{Naoki Aizawa} received the B. E. degree from The University of Electro-Communications, Tokyo, Japan, in 2024, and he is currently an M. E. student at The University of Electro-Communications, Tokyo, Japan.
His research interest includes encrypted controls.
\end{IEEEbiography}
\begin{IEEEbiography}[{\includegraphics[width=1in,height=1.25in,clip,keepaspectratio]{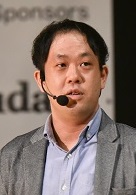}}]
{Keita Emura} received his M.E. degree from Kanazawa University in 2004. 
He was with Fujitsu Hokuriku Systems Ltd., from 2004 to 2006. He received his Ph.D. degree in information science from the Japan Advanced Institute of Science and Technology (JAIST) in 2010, where he was with the Center for Highly Dependable Embedded Systems Technology as a post-doctoral researcher in 2010-2012. 

He has been a researcher with the National Institute of Information and Communications Technology (NICT) since 2012, has been a senior researcher at NICT since 2014, and has been a research manager at NICT since 2021. 
His research interests include public-key cryptography and information security. 
He was a recipient of the SCIS Innovation Paper Award from IEICE in 2012, the CSS Best Paper Award from IPSJ in 2016, the IPSJ Yamashita SIG Research Award in 2017, and the Best Paper Award from ProvSec 2022. He is a member of IEICE, IPSJ, and IACR.
\end{IEEEbiography}
\begin{IEEEbiography}[{\includegraphics[width=1in,height=1.25in,clip,keepaspectratio]{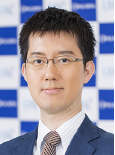}}]{Kiminao Kogiso} received his B.E., M.E., and Ph.D. degrees in mechanical engineering from Osaka University, Japan, in 1999, 2001, and 2004, respectively. 

He was appointed as a postdoctoral fellow in the 21st Century COE Program and as an Assistant Professor in the Graduate School of Information Science, Nara Institute of Science and Technology, Nara, Japan, in April 2004 and July 2005, respectively.
From November 2010 to December 2011, he was a visiting scholar at the Georgia Institute of Technology, Atlanta, GA, USA.
In March 2014, he was promoted to the position of Associate Professor in the Department of Mechanical and Intelligent Systems Engineering at The University of Electro-Communications, Tokyo, Japan. 
Since April 2023, he has been serving as a full Professor in the same department.
His research interests include cybersecurity of control systems, constrained control, control of decision-makers, and their applications.
\end{IEEEbiography}

\end{document}